\documentclass{optica-article}

\journal{opticajournal} 

\articletype{Research Article}

\usepackage{lineno}
\usepackage[numbers,compress]{natbib}
\usepackage{xcolor}
\usepackage{./styles/shortcuts}
\usepackage{amsmath}
\usepackage{gensymb}
\usepackage{tikzsymbols}
\usepackage{multirow}

\begin{document}

\title{Simple Method for Stripping Polyimide-Coated Optical Fiber}

\author{Matthew Marshall,\authormark{1,2} Jacob Williamson,\authormark{2,4} Seth Hyra,\authormark{4} Robert H. Leonard,\authormark{3} and Spencer E. Olson\authormark{4}}

\address{
\authormark{1}Department of Physics and Astronomy, University of New Mexico, Albuquerque, NM 87131\\
\authormark{2}Universities Space Research Association\\
\authormark{3} Space Dynamics Laboratory, Quantum Sensing \& Timing, North Logan, UT 84341, USA\\
\authormark{4}Air Force Research Laboratory, Kirtland Air Force Base, NM 87117, USA}

\email{\authormark{*}} 

\date{\today}

\newcommand{\distA}[1]{%
  Approved for public release; distribution is unlimited.  Public Affairs %
  release approval %
  \underline{#1}
}

\pagestyle{fancy}
\fancyhead{}
\renewcommand{\headrulewidth}{0pt}
\fancyfoot{}
\fancyfoot[R]{\thepage}
\fancyfoot[L]{
  \footnotesize
  \centering{\distA{AFRL-2025-4823}}
}


\begin{abstract*} 
We present a simple method for removing polyimide coatings from optical
fibers using inexpensive and readily available solvents. Impacts of solvent mixing
ratios, soak temperature, material expansion, wicking, and drying are
described to provide empirical context for the method. We find that soaking fibers for six hours in a 2:1 mixture of methanol to acetone at room temperature enables easy stripping of a length slightly greater than the soak length. 
\end{abstract*}


\section{Introduction}

\begin{figure}[htb]
    \centering
    \includegraphics[width=0.8\linewidth]{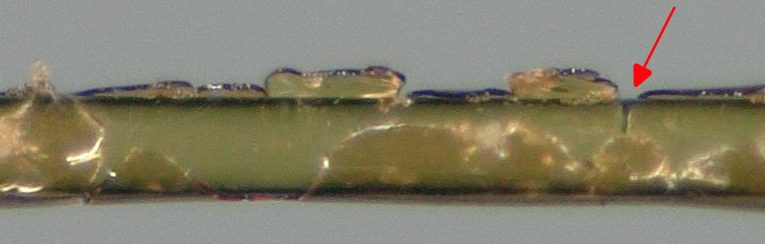}
    \caption{An example of shredding typically encountered when attempting to strip an untreated polyimide-coated fiber. Note the glass fiber has also already cracked where indicated by the red arrow.}
    \label{fig:shredding}
\end{figure}

Polyimide-coated fiber optics are often used in applications where the fiber
needs to be vacuum-compatible and well-protected from extreme
environmental conditions.  However, polyimide-coated fibers are difficult to strip, as typical fiber stripping techniques, such as simple thermal or mechanical stripping, are rendered ineffective.  For example, as shown in Figure~\ref{fig:shredding}, mechanical stripping tends to unevenly tear the coating, and frequently breaks the fiber.

The current standard methods to strip polyimide coatings are to
either (1) burn the coating off of the
fiber~\cite{duke_stripping_2011, cleveland_electric_laboratories_stripping_2017, song_study_2017, noauthor_vmm1000_nodate}, 
(2) strip the coating chemically with hot sulfuric
acid~\cite{duke_stripping_2011, noauthor_vmm1000_nodate}, or (3) use
a commercial machine for precision mechanical stripping~\cite{noauthor_thorlabs_nodate}.
These methods are expensive, dangerous, and/or difficult to perform without damaging the fiber.
The commercial options are often expensive enough to be impractical for many university labs~\cite{duke_stripping_2011, noauthor_thorlabs_nodate, noauthor_vmm1000_nodate}. In this paper,
we describe an alternative, simple method of chemically treating
and mechanically stripping polyimide coated fiber using chemicals
and tools commonly found in most labs working with optical fibers. 

\section{Description of Method}

Rather than utilizing expensive mechanical stripping gear or using a caustic
solution to breakdown or dissolve polyimide, we sought a method that was inexpensive, uses commonly-available solvents, and avoids weakening the fiber.  The new method we
present here involves preparing a section of polyimide-coated optical
fiber to be stripped by soaking it in a solvent mixture.  This soaking causes the polyimide coating to
expand and causes the fiber and polyimide to separate, allowing for
conventional strippers to remove the coating easily.

\begin{figure}[htb!]
    \centering
    \includegraphics[width=0.5\linewidth]{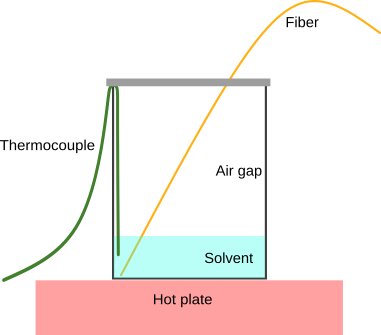}
    \caption{Fiber tips were soaked in a solvent mixture within a beaker. Any excess length of fiber passed through a pinhole in an aluminum foil covering to the beaker. The length of fiber submerged in the solvent and the length within the air gap are both much easier to strip after soaking.}
    \label{fig:fibersoak}
\end{figure}

Soaking the end of a polyimide-coated fiber in a 2:1 mixture of methanol and acetone for at least six hours at room temperature results in significantly increased ease of stripping using a standard model FTS4 three-hole stripping tool. Figure~\ref{fig:fibersoak} demonstrates the simple configuration used for the soaking.  The ease of stripping is most notable in the length of fiber that is either submerged in solution or directly exposed to the ambient vapors of the solution, respectively labeled as ``Solvent'' and ``Air gap'' in Figure~\ref{fig:fibersoak}.  The effect of the solvent is greatly reduced on the length of the fiber not directly exposed to the vapors. The required soak time can be reduced to roughly four hours when the soak temperature is increased to 40\degree~C. The fiber will dry out and become more difficult to strip after having been removed from the solvent for about ten minutes; however, this effect can be mitigated by dipping the fiber back into the solvent mixture.

\section{Testing Procedures}
\label{sec:Methods}

In order to explore our method, we sought to better understand the most obvious    
parameters that effect the solvent-mediated release, such as the solvent        
mixture, the duration of the soak, and the temperature of the solvent.
Additionally, we observed effects, such as changes in physical dimension and appearance, which we present here to aid in understanding the physical effects of exposing the fiber to liquid and vapor forms of the solvent.

To explore the parameter space, we soaked polyimide-coated fibers from ThorLabs (SM1550P) and
Fibercore (HB800P) in several mixtures of commonly-available solvents
for various lengths of time in a configuration as shown in Figure~\ref{fig:fibersoak}.  After soaking, the fibers were stripped using a standard model FTS4 three-hole stripping tool and the ease of stripping was subjectively compared to that of untreated fibers. This was tested both immediately after the soak, as well as after varied drying times. 
Additionally, we heated and varied the proportions of the most successful of these mixtures in an effort to further reduce soaking times.

Given that ``ease of stripping'' is an inherently subjective metric, we additionally measured visible changes to the fiber, namely the expansion of the coating both radially and axially, in an effort to quantify the effects of solvent soaks. A Keyence VHX-5000 digital microscope was used to characterize the expansion. Lastly, we examined the correlation between the length of fiber directly submerged in the solvent and the length of fiber which was easier to strip.


\section{Results and Discussion}

Table~\ref{tab:solvents} summarizes the results of experiments performed across various 1:1 solvent mixtures, wherein the ease of fiber stripping was subjectively characterized after soaking the fiber in the solvent for less than 12 hours.
From this table, we can see that pure methanol and a mixture of methanol and acetone produced the best results. Pure methanol moderately improved the ease of stripping, but only if the fiber was stripped within less than a minute after removal. However, soaking the fiber in a 1:1 solution of acetone and methanol rendered the fibers easy to strip for several minutes after removal from the solution, greatly reducing the time sensitivity for stripping as compared to pure methanol.
The ease of stripping was also significantly increased, and the soaked section of the coating could often be removed using two rubber fiber grippers: one gripper to hold the un-soaked portion of the fiber, while the other pulled the soaked coating from the fiber.  

\newcommand{\Yeah}{\Smiley[2][green!60!white]}
\newcommand{\Meh}{\Sey[2][yellow!60!white]}
\newcommand{\Nah}{\Sadey[2][red!60!white]}
\begin{table}[htb]
    \centering
    \begin{tabular}{c|c|c|c|}
         
         \cline{1-2}
         Methanol & \Meh \\ 
         \cline{1-3}
         Acetone & \Yeah & \Nah \\ 
         \hline
         Isopropanol & \Nah & \Nah & \Nah\\
         \hline
          & Methanol & Acetone & Isopropanol\\
    \end{tabular}
    \caption{
        All 1:1 mixtures of solvents tested. Each of the indicators highlight a
        subjective quality for how well the particular solvent mixture affected the polyimide fiber stripping.
    }
    \label{tab:solvents}
\end{table}

In an effort to improve our fiber stripping method, different proportions of the methanol and acetone mixture were tested.  Since methanol was the only single solvent we tested which was effective, we focused on increasing the methanol to acetone ratio.  Specifically, we tested 1:1, 2:1, 3:1, and 4:1 ratios of methanol to acetone. The 2:1 solution was consistently the most effective of the four. Best results were achieved after overnight soaks, although the same or similar ease of stripping could also be achieved after soaking for at least six hours at room temperature.

Another method we tested to reduce treatment time was increasing the temperature of the solvent.  We found this was effective at reducing the required soak time in both methanol and its mixtures with acetone. Due to the low boiling temperatures of these solvents, we heated the solution to between 40-50\degree C. Even with this constraint, the necessary soak time was cut by approximately one-third, with the 2:1 mixture requiring as little as four hours in the heated solvents to be effective.

The polyimide coating remains easily stripped with the 2:1 mixture for up to about ten minutes after removal from the solvent. However, changes in the polyimide allow it to take up the solvent again easily; Several fibers that had previously been soaked and dried were placed back in solution a month later. Only a ten-minute soak was required before the fibers could easily be stripped again.

\begin{figure}[htb]
    \centering
    \includegraphics[width=0.9\linewidth]{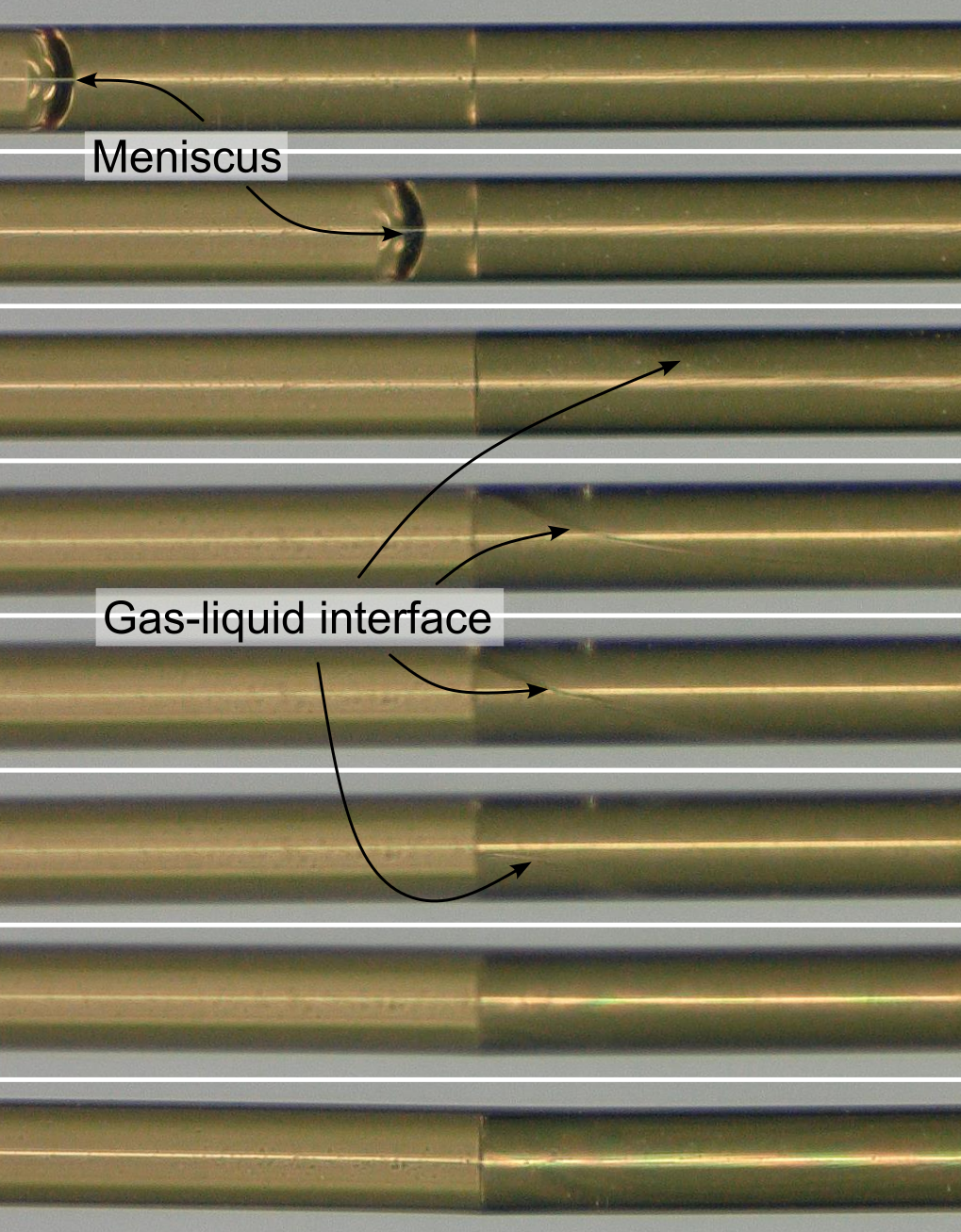}
    \caption{Examples of a visible layer of solvent between the polyimide coating and glass fiber. These images were taken in succession over the course of about 30 seconds. The glass fiber is on the right side of each image. The left side shows the distended length of the polyimide coating. Note the iridescence that becomes clear in the last two images.}
    \label{fig:thinfilm}
\end{figure}

While investigating the physical effects of soaking the fibers, we found that
measurements of radial expansion showed a maximal increase in diameter of about 6~$\mu\rm{m}$
. Additionally, there appeared to be a layer of solvent which formed between the polyimide coating and the glass fiber. When a fiber was placed under a microscope immediately after removal from solution, the solvent could be observed evaporating away as a meniscus in the distended portion of the coating that receded until it met the end of the glass fiber. Thereafter, lines of a gas-liquid interface could be observed as the layer of solvent between the fiber and the coating evaporated. After the solvent had fully evaporated, the fiber exhibited iridescent coloring, likely induced as a thin-film effect by the air remaining in the gap between glass and coating. This process takes about one minute to complete, and can be seen in Figure~\ref{fig:thinfilm}. Interestingly, after sufficient time to dry, the diameter of the fiber reverted to the nominal diameter or even up to 2~$\mu\rm{m}$ thinner than it had been prior to the soak. This contraction is a likely contributor to an observed increase in brittleness and decreased ease of stripping of fibers after drying. 


\begin{figure}[htb]
    \centering
    \includegraphics[width=0.9\linewidth]{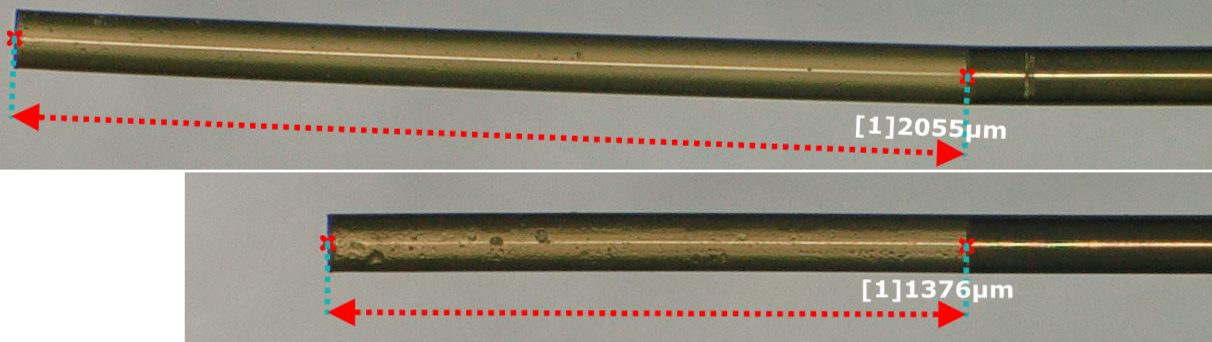}
    \caption{Two fibers pulled from the same 28-hour soak in 2:1 mixture. The upper image shows distention of about 2 mm, while the lower image shows distention below 1.4 mm, demonstrating the inconsistent nature of distention.}
    \label{fig:distension}
\end{figure}

In contrast, distension of the fiber coatings was readily apparent. After an overnight soak in either pure methanol or a mixed solution, fiber coatings frequently distended by several millimeters; however, the length of distention was inconsistent. Multiple fibers soaked for 25 hours in methanol before air-drying showed a total distension ranging from a few tens of microns to 1.4 cm in length. This distension retracts after drying by only roughly 2\% of the total distension. Two example fibers from these tests can be seen in Figure~\ref{fig:distension}.

Over the course of these tests, it was consistently noted that the easily-stripped length of fiber was longer than the length of fiber submerged in the solvents. To distinguish between the possible effects of capillary action drawing solvent further into the fiber and the coating interacting with solvent vapors, a fiber was isolated from vapor interactions by minimizing the air gap between the solvents and foil cover with only a few millimeters of fiber immersed in the solvents
. After a six-hour soak in 2:1 methanol-acetone solution, the length of fiber outside of the solution and air gap that had been reasonably affected by capillary action was approximately 6 cm. Over this length, an apparent gradient in ease of stripping was observed. The first 3 cm were stripped as easily as the end of a fiber that had soaked for roughly 3 hours at room-temperature, then further along it gradually became indistinguishable from an otherwise dry fiber.


Further tests used a large air gap to create an environment in which the suspected effects of capillary action and vapor interaction were combined. In these tests, a very sharp decrease in ease of stripping was found at the point at which the fiber protruded from the foil covering of the beaker. The length of fiber within the air gap was stripped with nearly identical ease to the submerged portions, suggesting that the effects of vapor permeation and capillary action on the coating is particularly significant when combined.

Drawing from these experiments, we suggest that the methanol is effective at permeating, softening, and expanding the polyimide coating.  This is further supported by measurements of methanol permeation in polyimide films found in the literature~\cite{musto_diffusion_2014, hausladen_permeation_1996}. Furthermore, we hypothesize that the methanol acts as a vehicle by which the acetone is introduced between the coating and fiber. Once the acetone reaches this space, it appears to dissolve the adhesive binding the coating to the fiber. This effect was evidenced by a residue left in the beakers after solvent evaporation, which appeared to have been a dissolved adhesive.  

\section{Conclusion}

We demonstrated that polyimide-coated fibers can be stripped with relative ease using a simple mixed-solvent bath. Soaking the fibers in a 2:1 methanol-to-acetone solution for a minimum of six hours at room temperature proved to be an approach that is highly effective, low risk, and is cheap and easy to implement. Furthermore, heating the mixture to approximately 40\degree~C effectively reduced the required soaking time to four hours, a decrease by one-third.  This provides a viable alternative to other approaches for polyimide-coated fiber preparation that are otherwise risky or expensive.

\section*{Disclaimer}
The views expressed are those of the authors and do not necessarily 
reflect the official policy or position of the Department of the Air 
Force, the Department of the Defense, or the U.S. Government. The authors declare no conflicts of interest. 


\begin{backmatter}
\bmsection{Funding}
This work was performed with funding from the Basic Research Office within the Office of the Under Secretary of Defense for Research and Engineering via the Laboratory University Collaboration Initiative.


\bmsection{Disclosures}
The authors declare no conflicts of interest.

\bmsection{Data Availability Statement} Data underlying the results presented in this paper are not publicly available at this time but may be obtained from the authors upon reasonable request.


\end{backmatter}


\bibliography{references}






\end{document}